\documentclass{article}%
\usepackage{amsmath}
\usepackage{amsfonts}
\usepackage{amssymb}
\usepackage{graphicx}%
\providecommand{\U}[1]{\protect\rule{.1in}{.1in}}

\begin{document}

\begin{center}
\textbf{Brans wormholes}

Kamal K. Nandi and Anwarul Islam

Department of Mathematics, University of North Bengal, Darjeeling (W.B.) 734
430 India

James Evans

Department of Physics, University of Puget Sound, Tacoma, Washington 98416

\textbf{Abstract}
\end{center}

It is shown that three of the four Brans solutions of classes I--IV admit
wormhole geometry. Two-way traversable wormholes in the Brans-Dicke theory are
allowed not only for the negative values of the coupling parameter $\omega$
($\omega<-2$), as concluded earlier, but also for arbitrary positive values of
$\omega$ ($\omega<\infty$). It also follows that the scalar field f plays the
role of exotic matter violating the weak energy condition.

\bigskip

PACS number(s): 04.20.Gz, 04.50.+h

\bigskip

Researches on wormhole physics by Morris, Thorne, and Yurtsever [1,2] have
opened up, in recent years, a new frontier in theoretical physics. There
already exist a number of investigations exploring the possible existence of
wormhole geometries in different physical situations [3--6]. The occurrence of
exotic matter having negative energy density [weak energy condition (WEC)
violation] offers an intriguing possibility as to whether wormholes might act
as effective gravitational lenses in astrophysical scenarios. Such a
possibility has been conjectured by Cramer \textit{et al}. [6] who also
recommend an analysis of massive compact halo objects (MACHO's) search data
for the detection of such lens effects. However, all the above analyses were
carried out only within the framework of Einstein's general relativity theory
(GRT). On the other hand, it is known that the GRT can be recovered in the
limiting case $\omega\rightarrow\infty$ of the Brans-Dicke theory (BDT). In
addition to the well-known utility of the BDT in local and cosmological
problems, it is often invoked in the interpretation of physical phenomena on a
galactic scale as well. For example, there are attempts aimed at explaining
the observed flat rotation curves in the vast domain of dark galactic haloes
[7,8]. It, therefore, seems only natural that in the context of wormhole
physics, too, one looks for wormhole solutions of BDT. The case of dynamic
wormholes has been dealt with by Accetta et al. [9] while the search for
static wormhole geometry in BDT has been initiated only recently by Agnese and
La Camera [10]. They show that a static spherically symmetric Brans-Dicke (BD)
solution, obtained in a certain gauge by Krori and Bhattacharjee [11], does
indeed support a two-way traversable wormhole for $\omega<-2$ and one way for
$\omega>-3/2$.

In the present paper, we wish to examine how many of the Brans I--IV classes
of solutions [12], which also include the case considered in [10], support
wormhole geometry. It is demonstrated that, of the four classes, as many as
three represent wormhole solutions provided the range of parameters are chosen
appropriately. The range, obtained by Agnese and La Camera [10], of the
coupling parameter for wormhole solutions, viz., $\omega<-2$, seems unduly
restrictive. Our analysis reveals that $\omega$ may take on arbitrary positive
values as well. It will also be apparent that the presence of the BD scalar
field f cannot prevent WEC violation showing that the latter is not a
consequence of the GRT alone.

The next four sections will deal with four classes of Brans solutions,
respectively. The final section concludes the results obtained in the paper.

The BD field equations are%

\begin{equation}
\square^{2}\varphi=\frac{8\pi}{3+2\omega}T_{M\mu}^{\mu},
\end{equation}

\begin{equation}
\mathbf{R}_{\mu\nu}-\frac{1}{2}g_{\mu\nu}\mathbf{R}=-\frac{8\pi}{\varphi
}T_{M\mu\rho}-\frac{\omega}{\varphi^{2}}\left[  \varphi_{,\mu}\varphi_{,\nu
}-\frac{1}{2}g_{\mu\nu}\varphi_{,\sigma}\varphi^{,\sigma}\right]  -\frac
{1}{\varphi}\left[  \varphi_{;\mu}\varphi_{;\nu}-g_{\mu\nu}\square^{2}%
\varphi\right]  ,
\end{equation}
where where $\square^{2}\equiv(\varphi^{;\rho})_{;\rho}$ and $T_{M\mu\rho}$ is
the matter energy-momentum tensor excluding the f field, $\omega$ is a
dimensionless coupling parameter. Brans [12] presented four classes of
solutions to BDT. The general metric, in isotropic coordinates ($t,r,\theta
,\varphi$) is given by ($G=c=1$)%
\begin{equation}
d\tau^{2}=-e^{2\alpha(r)}dt^{2}+e^{2\beta(r)}dr^{2}+e^{2\nu(r)}r^{2}%
(d\theta^{2}+\sin^{2}\theta d\psi^{2}).
\end{equation}
Brans solutions correspond to the gauge $\beta-\nu=0$. Class I solutions are
given by%
\begin{equation}
e^{\alpha(r)}=e^{\alpha_{0}}\left[  \frac{1-B/r}{1+B/r}\right]  ^{\frac
{1}{\lambda}},
\end{equation}%
\begin{equation}
e^{\beta(r)}=e^{\beta_{0}}\left[  1+B/r\right]  ^{2}\left[  \frac
{1-B/r}{1+B/r}\right]  ^{\frac{\lambda-C-1}{\lambda}},
\end{equation}%
\begin{equation}
\varphi(r)=\varphi_{0}\left[  \frac{1-B/r}{1+B/r}\right]  ^{\frac{C}{\lambda}%
},
\end{equation}%
\begin{equation}
\lambda^{2}\equiv(C+1)^{2}-C\left(  1-\frac{\omega C}{2}\right)  >0,
\end{equation}
where $\alpha_{0}$, $\beta_{0}$, $B$, $C$, and $\varphi_{0}$ are constants.
The constants $\alpha_{0}$ and $\beta_{0}$ are determined by asymptotic
flatness condition as $\alpha_{0}=$ $\beta_{0}=0$, while $B$ is determined by
the requirement of having Schwarzschild geometry in the weak field limit such
that $B=\lambda M/2$, $M>0$ is the central mass of the configuration. Clearly
$B$ and $\lambda$ must have the same sign.

The class I solution above is exactly the one considered in [10]. It can be
easily verified that Eq. (6) of [10] is just our Eq. (7) above. The important
point is that the exponents in Eqs. (4)--(6) depend on two parameters $\omega$
and $C$ satisfying the inequality (7). This implies that the range of $\omega$
is dictated by the range of $C$, which, in turn, is to be dictated by the
requirements of wormhole geometry as we shall see soon.

In their analysis, Agnese and La Camera [10] use post-Newtonian values to
parametrize their two exponents $A$ and $B$ (equivalently, our $\omega$ and
$C$) by a single parameter $\gamma\lbrack=(1+\omega)/(2+\omega)]$. This
procedure leads, after suitable readjustment of notations, to the equality
that $C=\gamma-1=-1/(\omega+2)$, which certainly constitutes a stronger
condition than the inequality [7]. As a further consequence, we find
$\lambda^{2}=(\omega+1.5)/(\omega+2)>0$, which implies that the range
$-2<\omega<1.5$ must be excluded a priori as it corresponds to imaginary
$\lambda$. Therefore, it seems more logical to use the inequality [7]
\textit{per se} for the analysis.

In order to investigate whether a given solution represents a wormwhole
geometry, it is convenient to cast the metric into Morris-Thorne canonical
form:%
\begin{equation}
d\tau^{2}=-e^{2\Phi(R)}dt^{2}+\left[  1-\frac{b(R)}{R}\right]  ^{-1}%
dR^{2}+R^{2}(d\theta^{2}+\sin^{2}\theta d\psi^{2}),
\end{equation}
where $\Phi(R)$ and $b(R)$ are called the redshift and shape functions,
respectively. These functions are required to satisfy some constraints,
enumerated in [1], in order that they represent a wormhole. It is, however,
important to stress that the choice of coordinates (Morris-Thorne) is purely a
matter of convenience and not a physical necessity. For instance, one could
equally well work directly with isotropic coordinates using the analyses of
Visser [3] but the final conclusions would be the same. Redefining the radial
coordinate $r\rightarrow R$ as%
\begin{equation}
R=re^{\beta_{0}}\left[  1+B/r\right]  ^{2}\left[  \frac{1-B/r}{1+B/r}\right]
^{\Omega}\text{, \ \ \ \ \ \ \ \ \ }\Omega=1-\frac{C+1}{\lambda},
\end{equation}
we obtain the following functions for $\Phi(R)$ and $b(R)$:%
\begin{equation}
\Phi(R)=\alpha_{0}+\frac{1}{\lambda}\left[  \ln\left\{  1-\frac{B}%
{r(R)}\right\}  -\ln\left\{  1+\frac{B}{r(R)}\right\}  \right]  ,
\end{equation}%
\begin{equation}
b(R)=R\left[  1-\left\{  \frac{\lambda\{r^{2}(R)+B^{2}\}-2r(R)B(C+1)}%
{\lambda\{r^{2}(R)+B^{2}\}}\right\}  ^{2}\right]  .
\end{equation}
The throat of the wormhole occurs at $R=R_{0}$ such that $b(R_{0})=R_{0}$.
This gives minimum allowed $r$-coordinate radii $r_{0}^{\pm}$ as%
\begin{equation}
r_{0}^{\pm}=B[(1-\Omega)\pm\sqrt{\Omega(\Omega-2)}]
\end{equation}
The values $R_{0}^{\pm}$ can be obtained from Eq. (10) using this $r_{0}^{\pm
}$. Noting that $R\rightarrow\infty$ as $r\rightarrow\infty$, we find that
$b(R)/R\rightarrow0$ as $R\rightarrow\infty$. Also, $b(R)/R\leq1$ for all
$R\geq$ $R_{0}^{\pm}$. The redshift function $\Phi(R)$ has a singularity at
$r=r_{S}=B$. In order that a wormhole be two-way traversable, the minimum
allowed values $r_{0}^{\pm}$ must exceed $r_{S}=B$. The extent to which this
requirement is satisfied depends on specific values of $\Omega$. Several cases
are possible.

(i) $-\infty<\Omega<0$ [$\Rightarrow\lambda<C+1$]. We see that $r_{0}^{+}>B$
while $r_{0}^{-}<B$. Hence a real, positive throat radius $R_{0}^{+}$ exists
only when $r=r_{0}^{+}$. The function $\Phi(R)$ is also nonsingular for $R\geq
R_{0}^{+}>0$ and it is finite everywhere. We therefore have a two-way
traversable wormhole. On the other hand, if $r=r_{0}^{-}$, the corresponding
value $R_{0}^{-}$ is imaginary and hence does not represent a wormhole.

(ii) $\Omega=0$ [$\Rightarrow\lambda=C+1$]. This gives a minimum allowed
radius $R_{0}^{\pm}=B$ and the function $\Phi(R)$ is singular at the
corresponding radius $R_{0}^{\pm}=4B$. Thus we obtain a non-Schwarzschild
one-way wormhole since $C\neq0$ and the scalar field $\phi$ is present. The
choice $C=0$ indicates the absence of the $\phi$ field and we have what is
known as the one-way Schwarzschild wormhole.

(iii) $0<\Omega<2$ [$\Rightarrow\lambda>C+1$]. In this case, $r_{0}^{\pm}$ and
hence $R_{0}^{\pm}$ are imaginary. Hence, no wormhole can be constructed.

(iv) $2\leq\Omega<\infty$. If $\lambda$\ assumes a positive sign and so does
$B$, then $r_{0}^{\pm}$ and $R_{0}^{\pm}$ both become negative and hence
wormholes are not possible. Let $\lambda$ assume a negative sign so that
$B=-B^{\prime}$,$B^{\prime}>0$. Then, from Eq. (12), we get $r_{0}%
^{-}>B^{\prime}$, $r_{0}^{+}<B^{\prime}$. The function $\Phi$ has no horizon
at $r=r_{0}^{-}$ and is finite for $r\geq r_{0}^{-}$ and we have a two-way
wormhole with a corresponding throat radius $R=R_{0}^{-}$. But if $r=r_{0}%
^{+}$, then $\Phi(R)$ is undefined, and we cannot have a wormhole. The case
$\Omega=2$ corresponds to case (ii) above.

Summing up, we see that two-way wormhole solutions are allowed only in the
ranges $0<\Omega<0$ and $2<\Omega<\infty$ (with $\lambda$ negative,
$\lambda=\lambda^{\prime}$. Let us write out $\Omega$ in terms of $\omega$ and
$C$\ explicitly:%
\begin{equation}
\Omega=1-\frac{C+1}{\lambda}=1-\frac{C+1}{\pm\left[  (C+1)^{2}-C\left(
1-\frac{\omega C}{2}\right)  \right]  ^{1/2}}.
\end{equation}
It is evident that $C+1$ and $\lambda$ must have the same sign for $\Omega<0$.
Suppose both have minus signs. Then, $C+1=-t,t>0$, say. The following
inequality must hold:%
\[
t>[t^{2}+(1+t)\{1+(\omega C/2)(1+t)\}]^{1/2}\Rightarrow(1+t)\omega<-2.
\]
It is possible to choose $t$ in such a way that v may take on any arbitrary
value in the open interval ($-2,0$). Suppose again that both $C+1$ and
$\lambda$ have plus signs. Then, $C+1=s,s>0$, say. The following must hold:%
\begin{align*}
s &  >[s^{2}-(s-1)\{1-(\omega C/2)(s-1)\}]^{1/2}\\
&  \Rightarrow-(s-1)\{1-(\omega C/2)(s-1)\}<0
\end{align*}
Now, two cases are possible: (a) If $0<s<1$, take $s-1=a$, then $a<0$. We then
have $a\omega<-2\Rightarrow-\infty<\omega<\infty$. (b) If $1<s<\infty$, take
$s-1=b>0$. Then, $b\omega<2$. In the limit $b\rightarrow0+$, we have
$\omega<\infty$. In other words, $\omega$ can take on arbitrary positive
values if $a$ and $b$ are appropriately chosen. For $2<\Omega<\infty$, we must
have ($C+1$)$>\lambda^{\prime}$ and we find $\omega<\infty$ from the same
analysis as above.

The combined energy density of the gravitational (second-order derivatives of
$g_{\mu\nu}$) + scalar ($\phi$) field $(T_{g}+T_{\phi})_{00}$ is obtained by
computing the Einstein tensor $G_{00}$ such that%
\begin{equation}
G_{00}=\frac{1}{8\phi}(T_{g}+T_{\phi})_{00}=\frac{1}{R^{2}}\frac{db}{dR}.
\end{equation}
From Eq. (11), we obtain%
\begin{equation}
\frac{db}{dR}=\frac{4r^{2}B^{2}}{(r^{2}-B^{2})^{2}}[\Omega(2-\Omega)].
\end{equation}
If $\Omega<0$ or $\Omega>2$, then $\frac{db}{dR}<0$. This implies that, with
$\phi$ everywhere non-negative, $G_{00}<0$. This shows that the scalar field
$\phi$\ plays the role of exotic matter at the wormhole throat. The same
conclusion was reached also in [10].

The axially symmetric embedded surface $z=z(R)$ shaping the wormhole's spatial
geometry is obtained from%
\begin{equation}
\frac{dz}{dR}=\pm\left[  \frac{R}{b(R)}-1\right]  ^{-1/2}.
\end{equation}
For a coordinate-independent description of wormhole physics, one may use
proper length $\ell$ instead of $R$ such that%
\begin{equation}
\ell=\pm\int_{R_{0}^{+}}^{R}\frac{dR}{[1-b(R)/R]^{1/2}}.
\end{equation}
In the present case,%
\begin{equation}
\ell=\pm\int_{r_{0}^{+}}^{r}e^{\beta(r)}dr.
\end{equation}
This integral is not integrable in a closed form. Nonetheless, it can be seen
that $\ell\rightarrow\pm\infty$ as $r\rightarrow\pm\infty$.

Class II solutions are given by%
\begin{equation}
\alpha(r)=\alpha_{0}+\frac{2}{\Lambda}\arctan\left(  \frac{r}{B}\right)  ,
\end{equation}%
\begin{equation}
\beta(r)=\beta_{0}-\frac{2(C+1)}{\Lambda}\arctan\left(  \frac{r}{B}\right)
-\ln\left(  \frac{r^{2}}{r^{2}+B^{2}}\right)  ,
\end{equation}%
\begin{equation}
\phi(r)=\phi_{0}e^{(2C/\Lambda)\arctan\left(  \frac{r}{B}\right)  },
\end{equation}%
\begin{equation}
\Lambda^{2}\equiv C\left(  1-\frac{\omega C}{2}\right)  -(C+1)^{2}.
\end{equation}
The constants $\alpha_{0}$ and $\beta_{0}$ are determined by using an
asymptotic flatness condition and the constant $B$ is determined by the weak
field condition as follows:%
\begin{equation}
\alpha_{0}=-\frac{\pi}{\Lambda}\text{, \ }\beta_{0}=\frac{\pi(C+1)}{\Lambda
}\text{, \ \ }B=\frac{\Lambda M}{2},
\end{equation}
where $M>0$ is the central mass of the configuration. The inequality (22)
fixes the range of $\omega$:$C\geq-1\Rightarrow\omega<-2$, or,
$C<-1\Rightarrow-2<\omega<-3/2$. The sign of $\Lambda$ is left undetermined.
Under the radial coordinate transformation $r\rightarrow R$%
\begin{equation}
r=r\left(  1+\frac{B^{2}}{r^{2}}\right)  \exp\left[  1-\frac{2}{\pi}%
\arctan\left(  \frac{r}{B}\right)  \right]  \beta_{0},
\end{equation}
class II solutions yield%
\begin{equation}
\Phi(R)=-\frac{\pi}{\Lambda}+\frac{2}{\pi}\arctan\left(  \frac{r(R)}%
{B}\right)  ,
\end{equation}%
\begin{equation}
b(R)=R\left[  1-\left\{  1+\frac{2B}{r^{2}(R)+B^{2}}\left(  \frac
{r(R)(C+1)}{\Lambda}-B\right)  \right\}  ^{2}\right]  .
\end{equation}
Once again, $R\rightarrow\infty$\ as $r\rightarrow\infty$\ and all the
conditions for a two way wormhole are satisfied by the above $\Phi(R)$ and
$b(R)$. The function $\Phi(R)$ has no horizon, is finite everywhere, and
$\Phi(R)\rightarrow\infty$ as $R\rightarrow\infty$. The $r$ radii of the
throat are given by%
\begin{equation}
r_{0}^{\pm}=\frac{B\beta_{0}}{\pi}[-1\pm(1+\beta_{0}^{2}/\pi^{2})^{1/2}].
\end{equation}
As usual, putting these values in Eq. (24), we can find $R_{0}^{\pm}$. Notice
that finite positive values of $r$ (except $r=0)$ correspond to finite
positive values of $R$. Thus we require that $r_{0}^{\pm}>0$ so that we can
have $R_{0}^{\pm}>0$. Rewriting Eq. (27) as $r_{0}^{+}=pM(1+C)$, where $p>0$
is any arbitrary real number, we find that the range $C>-1$ allows two-way
wormhole solutions since it ensures $r_{0}^{+}>0$. In the same way, $r_{0}%
^{-}=-qM(1+C)$, where $q>0$ is any arbitrary real number and $C<-1$ implies a
finite positive $R_{0}^{-}$ for the wormhole throat radius in the range
$-2<\omega<-3/2$.

It can be verified that%
\begin{equation}
\frac{db}{dR}\mid_{R=R_{0}^{\pm}}=-1
\end{equation}
and hence there occurs a WEC violation. The flaring-out condition
$d^{2}z/dR^{2}>0$ is also satisfied, since it can be verified that%
\begin{equation}
\frac{d^{2}z}{dR^{2}}\mid_{R=R_{0}^{\pm}}=\frac{1}{R_{0}^{2}}>0.
\end{equation}
The proper length $\ell$ is given by%
\begin{equation}
\ell=\pm e^{\beta_{0}}\int_{r_{0}^{+}}^{r}e^{\beta(r)}dr=\pm e^{\beta_{0}%
}[(r-r_{0}^{\pm})+...].
\end{equation}
Again, $R\rightarrow\pm\infty\Longleftrightarrow\ell\rightarrow\pm\infty$ as
$r\rightarrow\pm\infty$.

Class III solutions are given by%
\begin{equation}
\alpha(r)=\alpha_{0}-\frac{r}{B},
\end{equation}%
\begin{equation}
\beta(r)=\beta_{0}-\ln\left(  \frac{r}{B}\right)  ^{2}+(C+1)\left(  \frac
{r}{B}\right)  ,
\end{equation}%
\begin{equation}
\phi(r)=\phi_{0}e^{-(Cr/B)},
\end{equation}%
\begin{equation}
C=\frac{-1\pm\sqrt{-2\omega-3}}{\omega+2}.
\end{equation}
The redshift and shape functions are%
\begin{equation}
\Phi(R)=\alpha_{0}-\frac{r(R)}{B},
\end{equation}%
\begin{equation}
b(R)=R\left[  1-\left\{  1-\frac{C+1}{B}r(R)\right\}  ^{2}\right]  ,
\end{equation}
where%
\begin{equation}
R=r^{-1}B^{2}\exp\left(  \beta_{0}+\frac{C+1}{B}r\right)  .
\end{equation}
Here, too, $R\rightarrow\infty$\ as $r\rightarrow\infty$\ but
$b(R)/R\rightsquigarrow0$ as $R\rightarrow\infty$. Also $\Phi(R)\rightarrow
\infty$\ as $R\rightarrow\infty$. Asymptotic flatness condition is also not
satisfied by this solution. Therefore, there is no question of any wormhole
geometry in this case.

Class IV solutions are%
\begin{equation}
\alpha(r)=\alpha_{0}-\frac{1}{Br},
\end{equation}%
\begin{equation}
\beta(r)=\beta_{0}+\frac{C+1}{Br},
\end{equation}%
\begin{equation}
\phi(r)=\phi_{0}e^{-(Cr/B)},
\end{equation}%
\begin{equation}
C=\frac{-1\pm\sqrt{-2\omega-3}}{\omega+2}.
\end{equation}
Usual asymptotic flatness and weak field conditions fix $\alpha_{0}$,
$\beta_{0}$ and $B$ as%
\begin{equation}
\alpha_{0}=0\text{, \ }\beta_{0}=0\text{, \ \ }B=1/M>0.
\end{equation}
The functions are%
\begin{equation}
\Phi(R)=\alpha_{0}-\frac{1}{Br(R)},
\end{equation}%
\begin{equation}
b(R)=R\left[  1-\left\{  1-\frac{C+1}{Br(R)}\right\}  ^{2}\right]  ,
\end{equation}%
\begin{equation}
R=rB^{2}\exp\left(  \frac{C+1}{Br}\right)  .
\end{equation}
The wormhole throat occurs at%
\begin{equation}
r=r_{0}=\frac{C+1}{B}\Rightarrow R=R_{0}\left[  \frac{C+1}{B}\right]  e.
\end{equation}
It can be verified from Eq. (41) that $(C+1)>0$ only if $\omega<-2$. No
wormhole is possible if $-2<\omega\leq-3/2$ or $\omega>-3/2$, since $(C+1)$ is
either negative or imaginary.

The proper length is given by%
\begin{equation}
\ell=\pm\int_{r_{0}}^{r}\exp\left(  \frac{C+1}{Br}\right)  dr.
\end{equation}
One can see that if $r\rightarrow\pm\infty$, then $R\rightarrow\pm\infty$ and
$\ell\rightarrow\pm\infty$. It can be verified that all the conditions of a
two-way wormhole including the flaring-out condition are satisfied. The
peculiarity of this solution is that%
\begin{equation}
\frac{db}{dR}=-\left(  \frac{C+1}{Br}\right)  ^{2}<0
\end{equation}
and hence $G_{00}<0$ for all finite nonzero values of $r$ (and, of course,
$R$). This implies that the entire wormhole, and not only the throat, is made
up of exotic material.

The special case $C=-1$ is not of interest as it corresponds to a flat spatial section.

It was shown in the foregoing that three out of the four types of Brans
solutions give rise to a two-way traversable wormhole geometry provided the
constants are chosen appropriately. The restriction $\omega<-2$ need no longer
be strictly maintained, for, as we have seen, $\omega$ can also take on
positive values in the context of two-way wormholes. This result extends the
scope for the feasibility of wormhole scenarios even to the regime of ordinary
observations. For example, laser-ranging probes and observations on binary
systems put a lower limit of $\omega\geq500-600$ [13--15]. However, there
occurs a violation of the WEC at the wormhole throat even for $\omega<+\infty$
(class I solutions), but, unlike in [10], the range of $\omega$ (or $\gamma$)
alone does not cause it. The positive, real values of the throat radii
$r_{0}^{\pm}$ (or$R_{0}^{\pm}$) containing both $\omega$ and $C$ are actually
responsible for the WEC violation, as we have just seen. Only in class IV
solutions do we see that WEC is violated for all values of $r$.

A search for wormhole geometry in BDT amounts to an investigation of the
extent to which the scalar field $\phi$ does play the role of exotic matter
required for WEC violation. Researches into the existence of matter having
negative energy density (or, negative mass) are not new. It was Bondi [16] who
initiated the work and, in recent years, we have a number of investigations
into the question of negative energy [17--20]. Interestingly, Pollard and
Dunning-Davies [20] show that no contradictions arise if negative mass is
introduced into Newton's laws of motion.

\textbf{Acknowledgment}

One of us (A.I.) wishes to thank the Indian Council for Cultural Relations
(ICCR), Azad Bhawan, Indraprastha, New Delhi, for financial support through an
Exchange Program of the Government of India.

\textbf{References}

[1] M. S. Morris and K. S. Thorne, Am. J. Phys. \textbf{56}, 395 (1988).

[2] M. S. Morris, K. S. Thorne, and U. Yurtsever, Phys. Rev. Lett.
\textbf{61}, 1446 (1988).

[3] M. Visser, Phys. Rev. D \textbf{39}, 3182 (1989); Nucl. Phys.
\textbf{B328}, 203 (1989); \textit{Lorentzian Wormholes-From Einstein To
Hawking} (AIP, New York, 1995).

[4] V. P. Frolov and I. D. Novikov, Phys. Rev. D \textbf{42}, 1057 (1990).

[5] R. Balbinot, C. Barrabe`s, and A. Fabbri, Phys. Rev. D \textbf{51}, 2782
(1995); \textbf{49}, 2801 (1994).

[6] J. G. Cramer \textit{et al}., Phys. Rev. D \textbf{51}, 3117 (1995).

[7] N. Riazi and H. R. Askari, Mon. Not. R. Astron. Soc. \textbf{261}, 229 (1993).

[8] K. K. Nandi and A. Islam, Indian J. Phys. B \textbf{68}, 539 (1994).

[9] F. S. Accetta, A. Chodos, and B. Shao, Nucl. Phys. \textbf{B333}, 221 (1990).

[10] A. G. Agnese and M. La Camera, Phys. Rev. D \textbf{51}, 2011 (1995).

[11] K. D. Krori and D. R. Bhattacharjee, J. Math. Phys. \textbf{23}, 637 (1982).

[12] C. H. Brans, Phys. Rev. \textbf{125}, 2194 (1962).

[13] H. W. Zaglauer and C. M. Will (unpublished).

[14] J. V. Narlikar and A. K. Kembhavi, Fundam. Cosmic Phys. \textbf{6}, 1 (1980).

[15] J. V. Narlikar, \textit{Introduction to Cosmology }(Jones and Bartlett,
New York, 1983).

[16] H. Bondi, Rev. Mod. Phys. \textbf{29}, 423 (1957).

[17] W. B. Bonner, Gen. Relativ. Gravit. \textbf{21}, 1143 (1989).

[18] R. L. Forward, New Scientist \textbf{125}, 54 (1990).

[19] S. Kar, Phys. Rev. D \textbf{49}, 862 (1994).

[20] D. Pollard and J. Dunning-Davies, Nuovo Cimento B \textbf{110}, 857 (1995).

\end{document}